\documentclass[aps,preprint]{revtex4}%
\usepackage{amsfonts}
\usepackage{amsmath}
\usepackage{amssymb}
\usepackage{graphicx}%
\begin{document}
\title[ ]{ Quantum transport of Dirac electrons in graphene in the presence of a
spatially modulated magnetic field}
\preprint{ }
\author{M. Tahir$^{\ast}$}
\affiliation{Department of Physics, University of Sargodha, Sargodha 40100, Pakistan}
\author{K. Sabeeh}
\affiliation{Department of Physics, Quaid-i-Azam University, Islamabad 45320, Pakistan}
\author{}
\affiliation{}
\pacs{PACS number}

\begin{abstract}
We have investigated the electrical transport properties of Dirac electrons in
a monolayer graphene sheet in the presence of a perpendicular magnetic field
that is modulated weakly and periodically along one direction.We find that the
Landau levels broaden into bands and their width oscillates as a function of
the band index and the magnetic field.We determine the $\sigma_{yy}$ component
of the magnetoconductivity tensor for this system which is shown to exhibit
Weiss oscillations.We also determine analytically the asymptotic expressions
for $\sigma_{yy}$.We compare these results with recently obtained results for
electrically modulated graphene as well as those for magnetically modulated
conventional two-dimensional electron gas (2DEG) system.We find that in the
magnetically modulated graphene system cosidered in this work,Weiss
oscillations in $\sigma_{yy}$ have a reduced amplitude compared to the 2DEG
but are less damped by temperature while they have a higher amplitude than in
the electrically modulated graphene system. We also find that these
oscillations are out of phase by $\pi$ with those of the electrically
modulated system while they are in phase with those in the 2DEG system.

\end{abstract}
\volumeyear{year}
\volumenumber{number}
\issuenumber{number}
\eid{identifier}
\date[Date text]{date}
\received[Received text]{date}

\revised[Revised text]{date}

\accepted[Accepted text]{date}

\published[Published text]{date}

\startpage{1}
\endpage{2}
\maketitle

\section{\textbf{Introduction }}

The successful preparation of monolayer graphene has allowed the possibility
of studying the properties of electrons in graphene \cite{1}. The nature of
quasiparticles called Dirac electrons in these two-dimensional systems is very
different from those of the conventional two-dimensional electron gas (2DEG)
realized in semiconductor heterostructures. Graphene has a honeycomb lattice
of carbon atoms. The quasiparticles in graphene have a band structure in which
electron and hole bands touch at two points in the Brillouin zone. At these
Dirac points the quasiparticles obey the massless Dirac equation. In other
words, they behave as massless Dirac particles leading to a linear dispersion
relation $\epsilon_{k}=vk$ ( with the characteristic velocity $v\simeq
10^{6}m/s)$. This difference in the nature of the quasiparticles in graphene
from conventional 2DEG has given rise to a host of new and unusual phenomena
such as anamolous quantum Hall effects and a $\pi$ Berry phase\cite{1}%
\cite{2}. Earlier it was found that if conventional 2DEG is subjected to
artificially created periodic potentials in the submicrometer range it leads
to the appearence of Weiss oscillations in the magnetoresistance. This type of
electrical modulation of the 2D system can be carried out by depositing an
array of parallel metallic strips on the surface or through two interfering
laser beams \cite{3,4,5}.

Besides the fundamental interest in understanding the electronic properties of
graphene there is also serious suggestions that it can serve as the building
block for nanoelectronic devices \cite{6}. Since Dirac electrons can not be
confined by electrostatic potentials due to the Klein's paradox it was
suggested that magnetic confinement be considered \cite{7}. Technology for
this already exists as the required magnetic field can be created by having
ferromagnetic or superconducting layers beneath the substrate \cite{8}.

In conventional 2DEG systems, electron transport in the presence of magnetic
barriers and superlattices has continued to be an active area of research
\cite{9}. Recently, electrical transport in graphene in the presence of
electrical modulation was considered and theoretical predictions made
\cite{11}.Along the same lines, in this work we investigate low temperature
magnetotransport of Dirac electrons in a single graphene layer subjected to a
one-dimensional (1D) magnetic modulation. The perpendicular magnetic field is
modulated weakly and periodically along one direction

\section{Formulation and Energy Spectrum}

We consider two-dimensional Dirac electrons in graphene moving in the
x-y-plane. The magnetic filed ($B$) is applied along the z-direction
perpendicular to the graphene plane. The perpendicular magnetic field $B$ is
modulated weakly and periodically along one direction such that
$\overrightarrow{\mathbf{B}}=(B+B_{0}\cos(Kx))\widehat{z}$. Here $B_{0}$ is
the strength of the magnetic modulation. In this work we consider the
modulation to be weak such that $B_{0}<<B$. We consider the graphene layer
within the single electron approximation. The low energy excitations are
described by the two-dimensional (2D) Dirac like Hamiltonian ($\hbar=c=1$
here) \cite{1,2,11}%
\begin{equation}
H=v\overleftrightarrow{\sigma}.(-i\overrightarrow{\nabla}+e\overrightarrow
{A}). \label{1}%
\end{equation}
Here\ $\overleftrightarrow{\sigma}=\{\overleftrightarrow{\sigma}%
_{x},\overleftrightarrow{\sigma}_{y}\}$ are the Pauli matrices and $v$
characterizes the electron velocity. We employ the Landau gauge and write the
vector potential as $\overrightarrow{A}=(0,Bx+(B_{0}/K)\sin(Kx),0)$ where
$K=2\pi/a$ and $a$ is the period of the modulation. The Hamiltonian given by
Eq. (1) can be expressed as
\begin{equation}
H=-iv\overleftrightarrow{\sigma}.\overrightarrow{\nabla}+ev\overleftrightarrow
{\sigma_{y}}Bx+ev\overleftrightarrow{\sigma_{y}}\frac{B_{0}}{K}\sin(Kx).
\end{equation}
The above Hamiltonian can be written as%
\begin{equation}
H=H_{0}+H_{,}^{\prime} \label{2}%
\end{equation}
where $H_{0}$ is the unmodulated Hamiltonian given as
\[
H_{0}=-iv\overleftrightarrow{\sigma}.\overrightarrow{\nabla}%
+ev\overleftrightarrow{\sigma_{y}}Bx
\]
and
\[
H^{\prime}=ev\overleftrightarrow{\sigma_{y}}\frac{B_{0}}{K}\sin(Kx).
\]
The Landau level energy eigenvalues without modulation are given by%
\begin{equation}
\varepsilon(n)=\omega_{g}\sqrt{n} \label{4}%
\end{equation}
where $n$ is an integer and $\omega_{g}=v\sqrt{2eB}.$ As has been pointed out
\cite{11} the Landau level spectrum for Dirac electrons is significantly
different from the spectrum for electrons in conventional 2DEG\ which is given
as $\varepsilon(n)=\omega_{c}(n+1/2)$, where $\omega_{c}=eB/m$ is the
cyclotron frequency.

The eigenfunctions without modulation are given by
\begin{equation}
\Psi_{n,k_{y}}(r)=\frac{e^{ik_{y}y}}{\sqrt{2L_{y}l}}\left(
\begin{array}
[c]{c}%
-i\Phi_{n-1}[(x+x_{0})/l]\\
\Phi_{n}[(x+x_{0})/l]
\end{array}
\right)  \label{5}%
\end{equation}
where%
\begin{equation}
\Phi_{n}(x)=\frac{e^{-x^{2}/2}}{\sqrt{2^{n}n!\sqrt{\pi}}}H_{n}(x) \label{6}%
\end{equation}
where $l=\sqrt{1/eB}$ is the magnetic length, $x_{0}=l^{2}k_{y},$ $L_{y}$ is
the $y$-dimension of the graphene layer and $H_{n}(x)$ are the Hermite
polynomials. Since we are considering weak modulation $B_{0}<<$ $B$, we can
apply standard perturbation theory to determine the first order corrections to
the unmodulated energy eigenvalues in the presence of modulation%
\begin{equation}
\Delta\varepsilon_{_{n,k_{y}}}=%
{\displaystyle\int\limits_{-\infty}^{\infty}}
dx%
{\displaystyle\int\limits_{0}^{L_{y}}}
dy\Psi_{n,k_{y}}^{\ast}(r)H^{\prime}(x)\Psi_{n,k_{y}}(r) \label{7}%
\end{equation}
with the result%
\begin{equation}
\Delta\varepsilon_{_{n,k_{y}}}=\omega_{0}\cos(Kx_{0})\left(  2\sqrt{n}%
e^{-u/2}[L_{n-1}(u)-L_{n}(u)]\right)  \label{8}%
\end{equation}
where $\omega_{0}=\frac{evB_{0}}{K^{2}l}$, $u=K^{2}l^{2}/2$ and $L_{n}(u)$ are
the Laguerre polynomials. Hence the energy eigenvalues in the presence of
modulation are
\begin{equation}
\varepsilon(n,k_{y})=\varepsilon(n)+\Delta\varepsilon_{_{n,k_{y}}}=\omega
_{g}\sqrt{n}+\omega_{0}\cos(Kx_{0})G_{n} \label{9}%
\end{equation}
with\ $G_{n}(u)=2\sqrt{n}e^{-u/2}[L_{n-1}(u)-L_{n}(u)]$. We observe that the
degeneracy of the Landau level spectrum of the unmodulated system with respect
to $k_{y}$ is lifted in the presence of modulation with the explicit presence
of $k_{y}$ in $x_{0}.$ The $n=0$ landau level is different from the rest as
the energy of this level is zero and electrons in this level do not contribute
to diffusive conductivity calculated in the next section. The rest of the
Landau levels broaden into bands. The Landau bandwidths $\sim G_{n}$
oscillates as a function of $n$ since $L_{n}(u)$ are oscillatory functions of
the index $n$.

Before we begin the calculation of electrical conductivity it is necessary to
discuss the regime of validity of the perturbation theory presented above. For
large $n$ the level spacing given by Eq. (4) goes as $\omega_{g}(\sqrt
{n}-\sqrt{(n-1)})\longrightarrow\omega_{g}\frac{1}{2\sqrt{n}}$ and the width
of the $n$th level given by Eq.(8) goes as $2\omega_{0}n^{1/2}.$ There is
therefore a value of $n$ at which the width becomes equal to the spacing and
the perturbation theory is no longer valid. This occurs when $n_{\max}%
=\sqrt{2}\pi^{2}\frac{B^{\prime}}{B_{0}}$ where $B^{\prime}=\frac{1}{ea^{2}%
}=0.0054T$ for $a=350nm.$ For a fixed electron density and the period of
modulation this suggests the maximum value of the magnetic modulation $B_{0}$
above which it is necessary to carry out a more sophisticated analysis.

\section{Electrical Conductivity with Periodic Magnetic Modulation}

To calculate the electrical conductivity in the presence of weak magnetic
modulation we use Kubo formula to calculate the linear response to an applied
external field. In a magnetic field, the main contribution to Weiss
oscillations comes from the scattering induced migration of the Larmor circle
center. This is diffusive conductivity and we shall determine it following the
approach in \cite{10,11} where it was shown that the diagonal component of
conductivity $\sigma_{yy}$ can be calculated by the following expression in
the case of quasielastic scattering of electrons%

\begin{equation}
\sigma_{yy}=\frac{\beta e^{2}}{L_{x}L_{y}}\underset{\zeta}{%
{\displaystyle\sum}
}f(E_{\zeta})[1-f(E_{\zeta})]\tau(E_{\zeta})(\upsilon_{y}^{\zeta})^{2}
\label{10}%
\end{equation}
$L_{x}$, $L_{y}$, are the dimensions of the layer, $\beta=\frac{1}{k_{B}T}%
$\ is the inverse temperature with $k_{B}$ the Boltzmann constant, $f(E)$ is
the Fermi Dirac distribution function, $\tau(E)$\ is the electron relaxation
time and $\zeta$ denotes the quantum numbers of the electron eigenstate. The
diagonal component of the conductivity $\sigma_{yy}$ is due to modulation
induced broadening of Landau bands and hence it carries the effects of
modulation in which we are primarily interested in this work. $\sigma_{xx}$
does not contribute as the component of velocity in the $x$-direction is zero
here. The collisional contribution due to impurities is not taken into account
in this work.

The summation in Eq.(10) over the quantum numbers $\zeta$ can be written as%
\begin{equation}
\frac{1}{L_{x}L_{y}}\underset{\zeta}{%
{\displaystyle\sum}
}=\frac{1}{2\pi L_{x}}%
{\displaystyle\int\limits_{0}^{\frac{L_{x}}{l^{2}}}}
dk_{y}\underset{n=0}{\overset{\infty}{%
{\displaystyle\sum}
}}=\frac{1}{2\pi l^{2}}\underset{n=0}{\overset{\infty}{%
{\displaystyle\sum}
}} \label{11}%
\end{equation}
The component of velocity required in Eq.(10) can be calculated from the
following expression%
\begin{equation}
\upsilon_{y}^{\zeta}=\frac{\partial}{\partial k_{y}}\varepsilon(n,k_{y}).
\label{12}%
\end{equation}
Substituting the expression for $\varepsilon(n,k_{y})$ obtained in Eq.(9) into
Eq.(12) yields
\begin{equation}
\upsilon_{y}^{\zeta}=\frac{2\omega_{0}u}{K}\sin(Kx_{0})G_{n}(u) \label{13}%
\end{equation}
With the results obtained in Eqs.(11), (12) and (13) we can express the
diffusive contribution to the conductivity given by Eq.(10) as%
\begin{equation}
\sigma_{yy}=A_{0}\Phi\label{14}%
\end{equation}
where%
\begin{equation}
A_{0}=2\omega_{0}^{2}e^{2}\tau\beta\label{15}%
\end{equation}
and the dimensionless conductivity $\Phi$ is given as
\begin{equation}
\Phi=4ue^{-u}%
{\displaystyle\sum_{n=0}^{\infty}}
\frac{ng(E_{n})}{[g(E_{n})+1)]^{2}}[L_{n-1}(u)-L_{n}(u)]^{2} \label{16}%
\end{equation}
where $g(E)=\exp[\beta(E-E_{F}]$ and $E_{F}$ is the Fermi energy.

In Fig.(1), we plot the dimensionless conductivity given by Eq.(16) as a
function of inverse magnetic field at temperature $T=6K$ and electron density
$n_{e}=3\times10^{11}cm^{-2}$. The dimensionless magnetic field is introduced
given as $b=\frac{B}{B^{\prime}}$ with $B^{\prime}=\frac{1}{ea^{2}}.$In the
region of high magnetic field we can see SdH oscillations superimposed on the
Weiss oscillations.

\section{Asymptotic Expressions}

To get a better understanding of the results of the previous section we will
consider the asymptotic expression of conductivity where analytic results in
terms of elementary functions can be obtained following \cite{11}. We shall
compare the asymptotic results for the dimensionless conductivity obtained in
this section with the results obtained for a magnetically modulated
conventional 2DEG system. We shall also compare these results with those of
graphene that is subjected to only the electric modulation

The asymptotic expression of dimensionless conductivity can be obtained by
using the following asymptotic expression for the Laguerre polynomials%
\begin{equation}
\exp^{-u/2}L_{n}(u)\rightarrow\frac{1}{\sqrt{\pi\sqrt{nu}}}\cos(2\sqrt
{nu}-\frac{\pi}{4}). \label{17}%
\end{equation}
Note that the asymptotic results are valid when many Landau Levels are filled.
We \ now take the continuum limit:%
\begin{equation}
n-->\frac{1}{2}\left(  \frac{lE}{v}\right)  ^{2},\overset{\infty}%
{\underset{n=0}{%
{\displaystyle\sum}
}}-->\left(  \frac{l}{v}\right)  ^{2}%
{\displaystyle\int\limits_{0}^{\infty}}
EdE \label{18}%
\end{equation}
to express the dimensionless conductivity in Eq.(16) as the following integral%
\begin{equation}
\Phi=\frac{8\sqrt{u}}{\sqrt{2}\pi}\left(  \frac{l}{v}\right)  ^{3}%
{\displaystyle\int\limits_{0}^{\infty}}
dE\frac{E^{2}g(E)}{[g(E)+1)]^{2}}\sin^{2}(1/2\sqrt{u/n})\sin^{2}(2\sqrt
{nu}-\frac{\pi}{4}) \label{19}%
\end{equation}
where $u=2\pi^{2}/b.$

Now assuming that the temperature is low such that $\beta^{-1}\ll E_{F}$ and
replacing $E=E_{F}+s\beta^{-1}$, we rewrite the above integral as%
\begin{equation}
\Phi=\frac{8p^{2}a}{vb^{2}\beta}\sin^{2}\left(  \frac{\pi}{p}\right)
{\displaystyle\int\limits_{-\infty}^{\infty}}
\frac{dse^{s}}{(e^{s}+1)^{2}}\sin^{2}(\frac{2\pi p}{b}-\frac{\pi}{4}%
+\frac{2\pi a}{vb\beta}s) \label{20}%
\end{equation}
where $p=\frac{E_{F}a}{v}=k_{F}a=\sqrt{2\pi n_{e}}a$ is the dimensionless
Fermi momentum of the electron. To obtain an analytic solution we have also
replaced $E$ by $E_{F}$ in the above integral except in the sine term in the integrand.

The above expression can be expressed as
\begin{equation}
\Phi=\frac{8p^{2}a}{vb^{2}\beta}\sin^{2}\left(  \frac{\pi}{p}\right)
{\displaystyle\int\limits_{-\infty}^{\infty}}
\frac{ds}{\cosh^{2}(s/2)}\sin^{2}(\frac{2\pi p}{b}-\frac{\pi}{4}+\frac{2\pi
a}{vb\beta}s) \label{21}%
\end{equation}
The above integration can be performed by using the following identity
\cite{12}:%
\begin{equation}%
{\displaystyle\int\limits_{0}^{\infty}}
dx\frac{\cos ax}{\cosh^{2}\beta x}=\frac{a\pi}{2\beta^{2}\sinh(a\pi/2\beta)}
\label{22}%
\end{equation}
with the result%
\begin{equation}
\Phi=\frac{2p^{2}T}{\pi^{2}bT_{D}}\sin^{2}\left(  \frac{\pi}{p}\right)
\left[  1-A\left(  \frac{T}{T_{D}}\right)  +2A\left(  \frac{T}{T_{D}}\right)
\sin^{2}\left[  2\pi\left(  \frac{p}{b}-\frac{1}{8}\right)  \right]  \right]
\label{23}%
\end{equation}
where $k_{B}T_{D}=\frac{bv}{4\pi^{2}a},$ $\frac{T}{T_{D}}=\frac{4\pi^{2}%
a}{vb\beta}$and $A(x)=\frac{x}{\sinh(x)}-^{(x-->\infty)}->=2xe^{-x}.$

\section{Comparison with magnetically modulated conventional two-dimensional
electron gas system}

We start by comparing the energy spectrum and velocity expression obtained in
Eq.(9) and Eq.(13) with similar expressions for the conventional 2DEG where
the electron spectrum is parabolic \cite{9}. For the energy spectrum, we find
that the Landau level spectrum is significantly different from that of
standard electrons in conventional 2DEG. The first term $\omega_{g}\sqrt{n}$
with $\omega_{g}=v\sqrt{2eB}$ in Eq.(9) has to be compared with $\omega
_{c}(n+1/2)$ with $\omega_{c}=eB/m$ for standard electrons. The modulation
effects are in the second term where the essential difference is in the
structure of the function $G_{n}(u)=2\sqrt{n}e^{-u/2}[L_{n-1}(u)-L_{n}(u)].$
We find that there are essentially two basic differences: Firstly, for Dirac
electrons we have a difference of two successive Laguerre polynomials whereas
we had the sum of the Laguerre polynomials in the corresponding term for
standard electrons in 2DEG. Secondly, the expression for Dirac electrons is
multiplied by the square root of the Landau band index $\sqrt{n}$ that was
absent in the expression for standard electrons. The above mentioned
differences in the $G_{n}(u)$ function cause the velocity expression for the
Dirac electrons given by Eq.(13) to be different from that of the standard electrons.

As expected, these differences in the energy spectra and velocities lead to
different results for the diffusive conductivity in the two cases. We now
compare the results obtained for the asymptotic expression of the diffusive
conductivity $\sigma_{yy}$. To make this comparison possible we first express
$\Phi$ given by Eq.(23) as
\begin{equation}
\Phi=\frac{4k_{F}^{2}k_{B}T}{ve^{2}B^{2}a}\sin^{2}\left(  \frac{\pi}%
{p}\right)  F \label{24}%
\end{equation}
where $F=\left[  1-A\left(  \frac{T}{T_{D}}\right)  +2A\left(  \frac{T}{T_{D}%
}\right)  \sin^{2}\left[  2\pi\left(  \frac{p}{b}-\frac{1}{8}\right)  \right]
\right]  $. We now compare the results for dimensionless conductivity obtained
in Eq.(24) with those presented in Eq. (18) of \cite{9}(a). We find that the
result for graphene (Dirac electrons) system differs from that of the
conventional 2DEG (standard electrons) system by a factor $\frac{2p^{2}}%
{\pi^{2}b}\sin^{2}\left(  \frac{\pi}{p}\right)  .$ For the system under
consideration $p\thicksim50$ and in this limit if we take $\sin^{2}\left(
\frac{\pi}{p}\right)  \rightarrow\left(  \frac{\pi}{p}\right)  ^{2}$ it yields
the factor $\frac{2}{b}.$ Hence we conclude that the amplitude of the
oscillations in the conductivity will be reduced by this factor in the
magnetically modulated graphene system compared to the conventional 2DEG under
the same conditions. For the parameters considered in this work, the
conductivity is larger by a factor of $\approx44$ (at magnetic field $0.5T$)
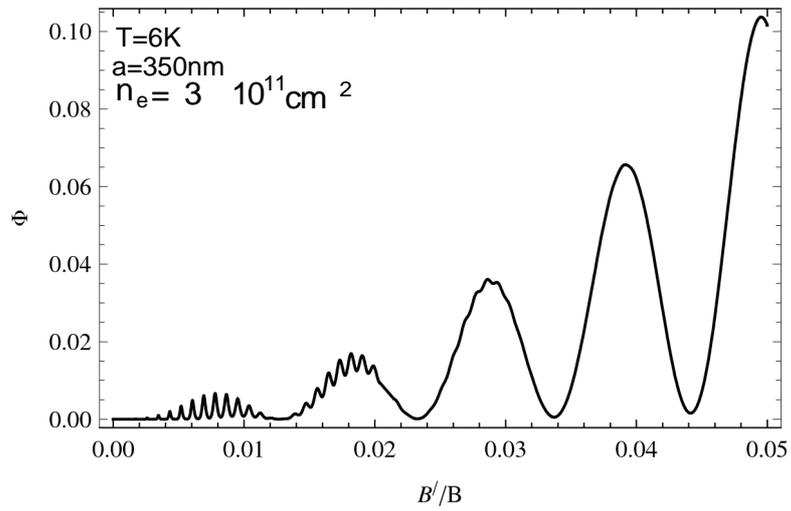
in the 2DEG system compared to graphene. In Fig.(2) we plot the dimensionless
conductivity versus inverse magnetic field for magnetically modulated graphene
and 2DEG. Note that in Fig.(2) the dimensionless conductivity for conventional
2DEG is rescaled by a factor .023.

The temperature scale for damping of Weiss oscillations in graphene can be
obtained from Eq.(23) and is characterized by $T_{D}$ given above while the
characteristic tempererature for 2DEG is given in \cite{9}(a) as $k_{B}%
T_{a}=\left(  \hbar\omega_{c}/4\pi^{2}\right)  ak_{F}.$ Comparing $T_{D}$ and
$T_{a}$ we obtain $\frac{T_{a}}{T_{D}}=\frac{v_{F}}{v}$ where $v_{F}$ is the
Fermi velocity in 2DEG. This shows that Weiss oscillations in graphene are
less damped with temperature compared to 2DEG due to the difference in Fermi
velocities in the two systems.

\section{Comparison with electrically modulated graphene system}

We will now compare the results obtained in this work with results obtained in
\cite{11} for the case of electrically modulated graphene system. We will
first compare the energy spectrum in the two cases. The difference in the
energy spectrum due to modulation effects was obtained in Eq.(8). If we
compare this result with the corresponding expression for the electrically
modulated case, we find the following differences: Firstly, in the magnetic
modulation case we have a difference of two successive Laguerre polynomials
whereas we had the average of two successive Laguerre polynomials in the
electric case. Secondly, in the magnetic modulation case the energy
eigenvalues are multiplied by the square root of the Landau band index
$\sqrt{n}$ that was absent in the expression for the electric case. These
differences cause the velocity expression for the Dirac electrons given by
Eq.(13) to be different from that of electrons in the electrically modulated system.

We now compare the expressions for dimensionless conductivity $\Phi$ given by
Eq. (23) with the electrically modulated case (Eq.(22) in \cite{11}). We find
that in the magnetically modulated case we have $\sin^{2}x$ functions in place
of $\cos^{2}x$ functions for the electric case which results in oscillations
being out of phase in the two cases. We also find that the amplitude of the
oscillations in the magnetic case are larger by a factor of $\frac{8\pi^{2}%
}{b}$ compared to the electrically modulated case. For the parameters
considered in this work, the conductivity is larger by a factor of
$\approx1.2$ (at magnetic field $0.5T$) in the magnetically modulated graphene
system compared to the electrically modulated one.

Exact expression of the dimensionless conductivity $\Phi$ for the electric and
magnetic modulated graphene system is shown in Fig.(3) as a function of the
inverse magnetic field at temperature $T=6K$ , electron density $n_{e}%
=3\times10^{11}cm^{-2}$ and period of modulation $a=350nm$. In Fig.(3) we can
clearly see that the Weiss oscillations in the dimensionless conductivity are
enhanced, they have a larger amplitude, in the magnetically modulated case
compared to the electrically modulated case for the same parameter values.
Furthermore, we note that the oscillations in the magnetic and electric
modulated cases have a $\pi$ phase shift. We also observe that in the region
of high magnetic field SdH oscillations are superimposed on the Weiss
oscillations. The oscillations are periodic in $1/B$ and the period depends on
electron density as $\sqrt{n_{e}}$ in both the magnetic and electric modulated
cases. The characteristic damping temperature is the same for both the systems.

To better understand the increase in amplitude of Weiss oscillations in the
magnetically modulated graphene system compared to the electrically modulated
one we consider the difference in bandwidths in the two cases \cite{13}.
Important feature is the additional $\sqrt{n}$ factor in the perturbed energy
eigenvalues for the magnetically modulated case which is absent in the
electrically modulated case. The result is that the bandwidth in the
magnetically modulated case is approximately greater by a factor of $4\sqrt
{u}$ compared to the electrically modulated graphene system.

\section{Conclusions}

In this work we have investigated the electrical transport properties of Dirac
electrons in a monolayer graphene sheet in the presence of a perpendicular
magnetic field that is modulated weakly and periodically along one direction.
Our primary focus has been the study of Weiss oscillations in the diffusive
magnetoconductivity $\sigma_{yy}$ of this system. We have compared the results
obtained with those obtained for magnetically modulated conventional 2DEG
system and with those of the graphene system subjected to only the electric
modulation. We find that in the magnetically modulated graphene system Weiss
oscillations in the magnetoconductivity have a reduced amplitude compared to
the conventional 2DEG but are more robust with respect to temperature. In
comparison with the electrically modulated graphene case, we find that the
conductivity is larger in amplitude. We also find that the oscillations in the
magnetoconductivity in graphene are $\pi$ phase shifted with respect to the
electrically modulated case whereas they are in phase with the conventional
2DEG subjected to magnetic modulation.

\section{Acknowledgements}

One of us (K.S.) would like to acknowledge the support of the Pakistan Science
Foundation (PSF) through project No. C-QU/Phys (129). M. T. would like to
acknowledge the support of the Pakistan Higher Education Commission (HEC).

$\ast$Present address: Department of Physics, Blackett Laboratory, Imperial
College London, London SW7 2AZ, United Kingdom.

\end{document}